\documentclass[10pt,twocolumn,pra,aps,superscriptaddress]{revtex4-1}
\usepackage[latin9]{inputenc}
\usepackage{amsmath}
\usepackage{amssymb}
\usepackage{graphicx}


\begin{document}

\title{Light polarization measurements in tests of macrorealism}

\author{Eugenio Rold\'an}
\affiliation{Departament d'\`Optica i d'Optometria i Ci\`encies de la Visi\'o, Universitat de Val\`encia, Dr. Moliner 50, 46100 Burjassot, Spain.}
\affiliation{Max-Planck-Institut f\"ur Quantenoptik, Hans-Kopfermann-strasse 1, 85748 Garching, Germany.}

\author{Johannes Kofler}
\affiliation{Max-Planck-Institut f\"ur Quantenoptik, Hans-Kopfermann-strasse 1, 85748 Garching, Germany.}

\author{Carlos Navarrete-Benlloch}
\affiliation{Max-Planck-Institut f\"ur Quantenoptik, Hans-Kopfermann-strasse 1, 85748 Garching, Germany.}
\affiliation{Max-Planck-Institut f\"ur die Physik des Lichts, Staudtstrasse 2, 91058 Erlangen, Germany.}
\affiliation{Institute for Theoretical Physics, Erlangen-N\"urnberg Universit\"at, Staudtstrasse 7, 91058 Erlangen, Germany.}

\begin{abstract}
According to the world view of macrorealism, the properties of a given
system exist prior to and independent of measurement, which is incompatible
with quantum mechanics. Leggett and Garg put forward a practical criterion
capable of identifying violations of macrorealism, and so far experiments
performed on microscopic and mesoscopic systems have always ruled
out in favor of quantum mechanics. However, a macrorealist can always
assign the cause of such violations to the perturbation that measurements
effect on such small systems, and hence a definitive test would require
using non-invasive measurements, preferably on macroscopic objects,
where such measurements seem more plausible. However, the generation
of truly macroscopic quantum superposition states capable of violating
macrorealism remains a big challenge. In this work we propose a setup
that makes use of measurements on the polarization of light, a property
which has been extensively manipulated both in classical and quantum
contexts, hence establishing the perfect link between the microscopic
and macroscopic worlds. In particular, we use Leggett-Garg inequalities
and the criterion of no-signaling in time to study the macrorealistic
character of light polarization for different kinds of measurements,
in particular with different degrees of coarse-graining. Our proposal
is non-invasive for coherent input states by construction. We show for states with well defined photon number in
two orthogonal polarization modes, that there always exists a way of making
the measurement sufficiently coarse-grained so that a violation of
macrorealism becomes arbitrarily small, while sufficiently sharp measurements
can always lead to a significant violation. 
\end{abstract}
\maketitle

\section{Introduction}

In 1985, Leggett and Garg introduced the concept of macroscopic realism
(macrorealism) \cite{LG85,LGreview}. According to this worldview,
macroscopic objects are always in a state with well defined properties
and measurements can be performed without changing these properties
and the subsequent temporal evolution. If effects of decoherence can
be sufficiently suppressed, quantum mechanics is at variance with
macrorealism. While experiments with genuine macroscopic objects (``Schrödinger
cats'') have not been performed yet, the past few years have demonstrated
that quantum mechanics still prevails for mesoscopic objects \cite{Palacios10,Souza11,Knee12,Lvovsky13,Asadian14,Knee16}.

Leggett-Garg inequalities (LGIs) can witness a violation of macrorealism
by suitably comparing correlations of a dichotomized observable that
is measured at subsequent times under different measurement settings.
The concept of non-invasive measurability is essential for their derivation:
provided that the disturbance on the target system is negligibly small
during the measurement, the violation of a LGI would imply that the
corresponding observable does not obey macrorealism. Recently, an
alternative criterion for macrorealism called no-signaling in time
(NSIT) \cite{Kofler13} has been put forward, which is now known to
be in general stronger than LGIs \cite{Clemente16}, providing not
only sufficient but also necessary conditions for the absence of macrorealism
\cite{Clemente15}.

While the preparation of exotic quantum states such as macroscopic
spatial superpositions remains one of today's most anticipated challenges
in quantum mechanics \cite{Armour02,Schwab05,Arndt14,Abdi16,ORI16},
the polarization of light, with a long experimental history both in
classical and quantum physics, offers a perfect property where studying
macrorealism. Experiments have indeed been performed with single photons
in a superposition of two orthogonal polarization states\textbf{ }\cite{Goggin11},
following closely the original Leggett-Garg proposal. Although it
is widely accepted that the results of this study are in favor of
quantum mechanics, they seem to suggest that violations of macrorealism
weaken as the measurements are made less invasive, and the biggest
violations are found whenever the weak values are \emph{strange}.
In addition, violations of macrorealism can be traced back in this
case directly to the superposition of the two orthogonal polarizations.
In this work we consider a different setup, involving (strong) measurements
whose invasiveness can be varied in two different ways, and which
allow for violations of macrorealism for a broader type of states,
including some which make no apparent use of the superposition principle.

As explained in detail below, starting from a well defined spatiotemporal
mode of the light field, we propose using (polarization-insensitive)
beam splitters to perform photon counting measurements of the reflected
beam's polarization. Since coherent states remain coherent after the
beam splitter, the setup doesn't change the polarization state for
such states, and hence it can be regarded as non-invasive in the classical
limit. In addition, it is shown that the measurements preserve the
polarization state when the initial state is pure and has well-defined
polarization. It follows that a requirement for violations of macrorealism
in our setup is the presence of quantum states with a degree of polarization
smaller than one, which is a broader condition than having superpositions
of different polarization states.

For definiteness, we concentrate on linear polarization states with
well-defined photon number, even if some comments will be made about
other types of states. We pay special attention to the influence of
the type of measurements, as we consider measurements consisting in
the extraction of a fixed number of photons, but also measurements
that average over the number of detected photons. Our work can be
seen then as a study about the invasiveness of polarization measurements
via photon subtraction, and sheds light onto the question under which
measurement conditions a violation of macrorealism can be observed.
Our results are in agreement with earlier studies which show that,
for most time evolutions, sufficiently coarse-grained measurements
do not show a violation of macrorealism while sufficiently sharp measurements
do \cite{Kofler07,Wang13,Jeong14,Sekatski14,Clemente15}.

\begin{figure}
\includegraphics[width=1\columnwidth]{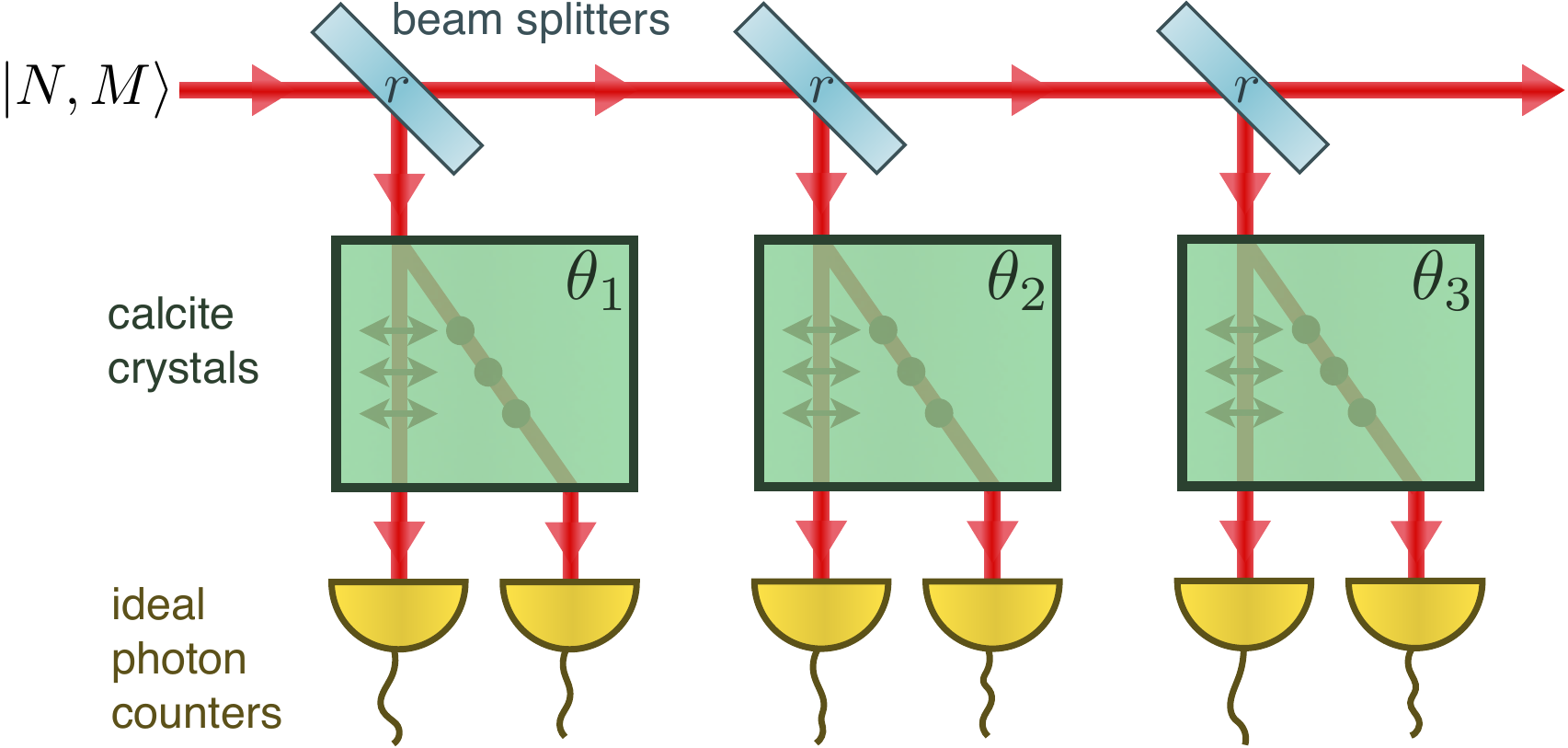}\caption{Sketch of the measurement protocol. Light pulses (following the red
path) are prepared in a state with well defined photon numbers in
the $x$ and $y$ linear polarizations. Each detection port acts following
the same three steps: (i) the pulses go through a beam splitter with
reflectivity $r$, (ii) their polarization components with respect
to some linear basis defined by an angle $\theta_{p}$ are separated
with a calcite crystal or any other method of choice, and (iii) finally
the photon number is measured on each polarization. Violations of
macrorealism can be then checked by studying the correlations between
measurements on the different ports. \label{Fig1}}
\end{figure}

\section{The protocol}

Consider a source of quantum light emitting a sequence of identical
pulses traveling along the $z$ direction, in a well defined transverse
spatial mode. Each pulse is prepared in a state with $N$ and $M$
photons linearly polarized along the $x$ and $y$ directions, respectively,
a state denoted by $\left|N,M\right\rangle $. The pulses travel through
three detection ports which act exactly in the same way, see Fig.
\ref{Fig1}. A fraction of the input light is extracted via a beam
splitter of (polarization-independent) reflectivity $r$, which will
be considered infinitesimally small when aiming for a non-invasive
measurement, although we will also study the effect of larger reflectivity
values. This small amount of reflected light is then split into its
orthogonal linear polarization components with respect to some $\theta$\textendash oriented
reference frame (this can be accomplished, e.g., by a properly oriented
calcite crystal), and each of the two orthogonally polarized beams
impinge on ideal photon counters that we denote by $x$ and $y$ \emph{detectors}.
We assume that the reflectivity $r$ is the same for all the detection
ports, while they might differ in the polarization angles $\theta$.
In order to study LGIs and NSIT, one must analyze correlations between
the statistics of the measurements at the different detection ports,
as we explain in detail below.

Note that each port allows performing a measurement of the polarization
state of the reflected beam, since accumulating the measurement results
pulse after pulse to determine the probability distributions at the
photon counters, one can unveil the statistics of the Stokes parameters
(when measuring along two different polarization orientations). It
is easy to show that the statistics of the Stokes parameters are the
same for the reflected and input beams for coherent and thermal states
(making the device a practical one for measuring the polarization
state of classical light), which is also true for Fock states at least
for first and second order moments.

\begin{figure*}[t]
\includegraphics[width=1\textwidth]{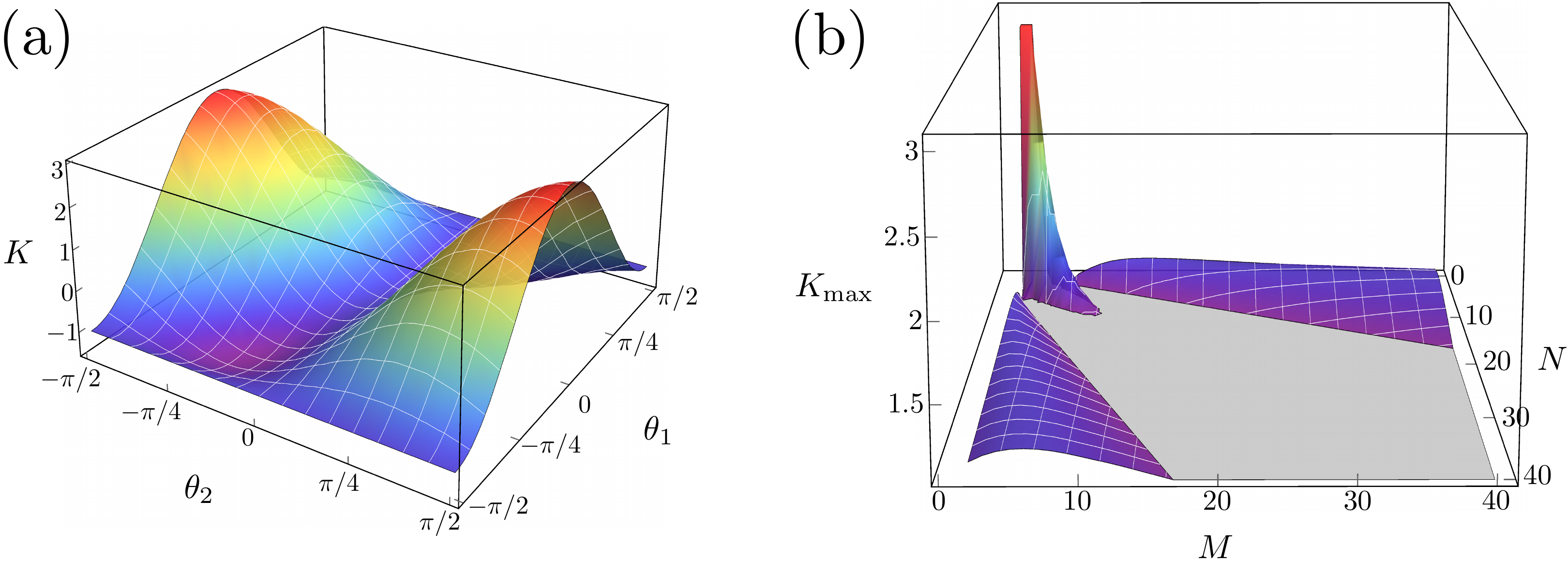}\caption{(a) $K$ as a function of $\theta_{2}$ and $\theta_{3}$ for $\theta_{1}=\pi/2$
and for the state $|1,1\rangle$ and $\mathcal{S}_{1}$ measurements.
For this type of measurements this is the only one that violates the
LGI, and it does so maximally, as $K_{\max}=3$. (b) $K_{\text{max}}$
as a function of $N$ and $M$ for $\mathcal{S}_{2}$ measurements.
The grey area marks the region where the LGI is preserved, and it
can be appreciated that the violation remains for large values of
$N$ and $M$. \label{Fig2}}
\end{figure*}

Let us now explain how do we dichotomize the measurement outcomes.
The detectors are capable of counting photons, and hence in principle
they will provide outcomes $\omega_{x}=0,1,2,...$ and $\omega_{y}=0,1,2,...$,
where the subindex labels the detector. We consider different strategies
to dichotomize these outcomes, all requiring post-selection and based
on some kind of ``majority vote\textquotedblright , that is, which
detector has detected more photons:

(a) $\mathcal{S}$ (sharp) measurements, in which one of the detectors
doesn't click, but the other measures some given number of photons
$\omega\geqslant1$ that we choose. Hence, we post-select to events
with either $(\omega_{x}=\omega,\omega_{y}=0)$ or $(\omega_{x}=0,\omega_{y}=\omega)$.
In the first case we say that the outcome corresponds to an $x$\emph{-event},
while it is a $y$\emph{-event} in the second case. We will use the
notation $\mathcal{S}_{\omega}$ whenever we want to refer explicitly
to the value of $\omega$ that we chose.

(b) $\mathcal{F}$ (fair) measurements, in which we keep all the outcomes
up to some maximum value $\omega_{\max}$ in both detectors. Any outcome
in which the $x$ detector has measured more photons than the $y$
detector is characterized as an $x$-event, and vice versa, outcomes
in which the $y$ detector has measured more photons than the $x$
detector are characterized as a $y$-event. As with the previous type,
we will use the notation $\mathcal{F}_{\omega_{\textrm{max}}}$ when
needed.

(c) $\mathcal{B}$ (blurred, or intermediate) measurements, in which
only outcomes from a certain photon-number interval $[\omega_{\min},\omega_{\max}]$
are considered, and the dichotomization is performed as in the previous
case. When in need of being more explicit, we will denote these measurements
by $\mathcal{B}_{[\omega_{\mathrm{min}},\omega_{\mathrm{max}}]}$.
Note that by taking $\omega_{\min}=1$ we recover type $\mathcal{F}_{\omega_{\mathrm{max}}}$,
while by taking $\omega_{\min}=\omega_{\max}=\omega$ we recover type
$\mathcal{S}_{\omega}$.

The above types of measurements affect the incoming pulse in different
ways. How invasive they are depends both on the beam splitter reflectivity
$r$, as it limits the number of photons that can be extracted from
the light beam, and on the choice of measurement, since these are
``selective'' in different degrees, in the sense that they make
use of more or less likely events. The least invasive measurement
would be one that is minimally reflecting and minimally selective,
that is, it makes use of all the photon counts. The ideal non-invasive
measurement corresponds then to an $\mathcal{F}_{\omega_{\textrm{max}}}$
measurement satisfying the condition $(N+M)r^{2}\ll1<\omega_{\mathrm{max}}$.
Contrarily, the $\mathcal{S}$ type is the more selective, while the
$\mathcal{B}$ type allows us to move in between the $\mathcal{F}$
and $\mathcal{S}$ types.

\section{Criteria for violations of macrorealism}

In order to evaluate both the LGIs and NSIT, it is necessary to evaluate
not only the probabilities $P_{a}(\theta)$ of having an $a$-event
for a setting $\theta$ of the detector (in the following the indices
$a$, $b$, and \textbf{$c$} take values from the event labels $x$
and $y$), but also the conditional probabilities $P_{ab}(\theta_{p},\theta_{q})$
of having an $a$-event at one device with polarization angle $\theta=\theta_{p}$
followed by a $b$-event at a subsequent device with $\theta=\theta_{q}$
(in the following, the indices $p$ and $q$ take values from the
measurement ports 1, 2, and 3), which we will denote as an $ab$-\emph{event}.
For NIST, we will also need to introduce $abc$-\emph{events}, with
associated probability $P_{abc}(\theta_{1},\theta_{2},\theta_{3})$,
related to three consecutive measurements performed over the same
pulse (with measurement $p$ at a polarization angle $\theta_{p}$).
The mathematical expressions for all these probabilities, as well
as their detailed derivation within the quantum mechanical framework,
can be found in the appendix. In the following we move directly to
introducing our criteria for violations of macrorealism and discussing
the results.

Let us now introduce the criteria based on LGIs and NIST. The basic
objects required to compute LGIs are the correlation functions between
two measurement ports 
\begin{equation}
C_{pq}\left(\theta_{p},\theta_{q},r\right)=\frac{P_{xx}+P_{yy}-P_{xy}-P_{yx}}{P_{xx}+P_{yy}+P_{xy}+P_{yx}},
\end{equation}
where we have included explicitly the dependence on the reflectivity
coefficient of the beam splitters. Together with the type of measurement,
and the number of photons $N$ and $M$ in the input state, these
are all the variables that define the problem. Note that whenever
$C_{pq}=+1\:(-1)$ the devices show perfectly correlated (anticorrelated)
results. The LGI reads then \cite{LGreview} 
\begin{equation}
K\left(\theta_{1},\theta_{2},\theta_{3},r\right)=C_{12}+C_{23}-C_{13}\leq1.
\end{equation}
A convenient witness is obtained by maximizing this quantity over
the polarization angles, defining then 
\[
K_{\max}\left(r\right)=\max_{\theta_{1},\theta_{2},\theta_{3}}K\left(\theta_{1},\theta_{2},\theta_{3},r\right),
\]
which witnesses a violation of the LGI (absence of macrorealism) whenever
it is larger than 1 (note that it is upper-bounded by 3).

\begin{figure*}[t]
\includegraphics[width=0.85\textwidth]{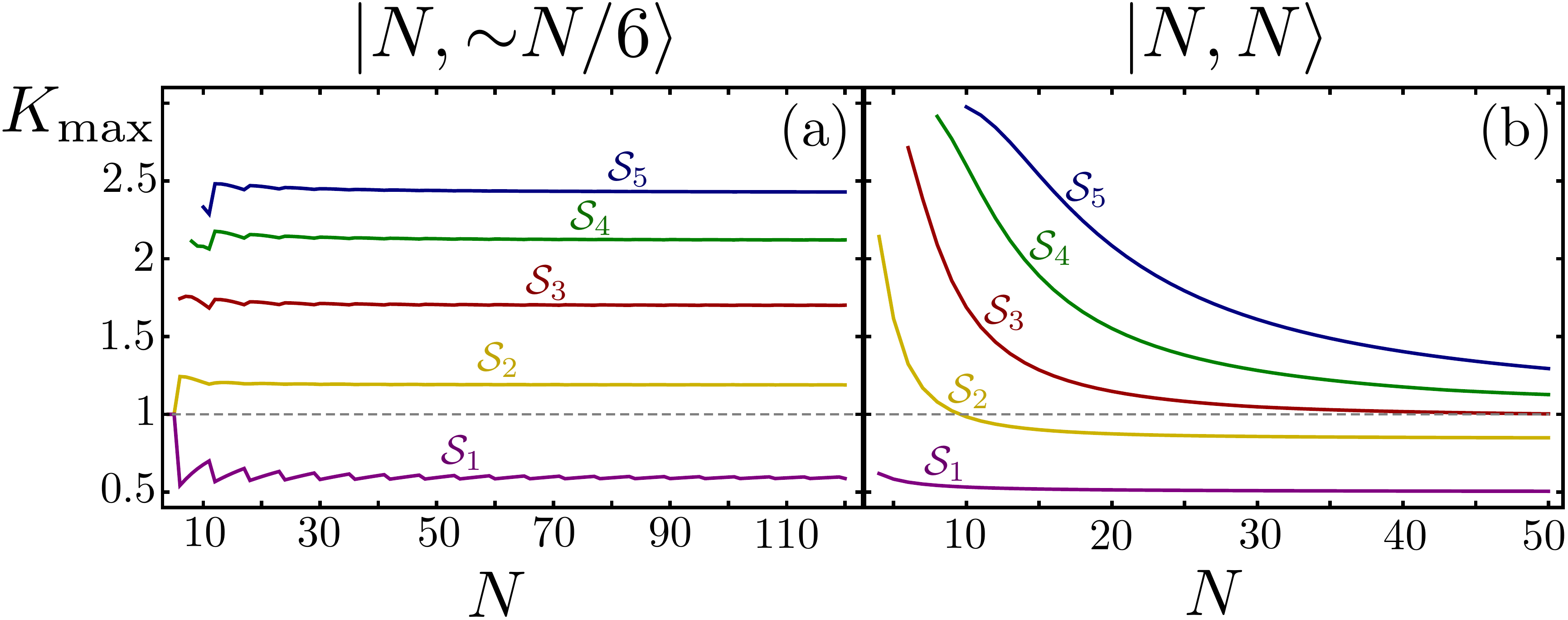}\caption{$K_{\max}$ for states $|N,\sim N/6\rangle$ and $|N,N\rangle$ in
(a) and (b), respectively, as a function of $N$ for different types
of $\mathcal{S}_{\omega}$ measurements (as indicated for each line).
\label{Fig4}}
\end{figure*}

NSIT requires analyzing the disturbance effected by one measurement
device on the others \cite{Kofler13}. When only two consecutive measurements
are considered, we then need to compare the probability distributions
on the second device without and with a measurement in a previous
device. These are given, respectively, by \begin{subequations} 
\begin{align}
\mathcal{P}_{b}\left(\theta_{2}\right) & =\frac{P_{b}\left(\theta_{2}\right)}{\sum_{a}P_{a}\left(\theta_{2}\right)},\\
\mathcal{P}_{b}^{'}\left(\theta_{1},\theta_{2}\right) & =\frac{\sum_{a}P_{ab}\left(\theta_{1},\theta_{2}\right)}{\sum_{a,b}P_{ab}\left(\theta_{1},\theta_{2}\right)},
\end{align}
\end{subequations} which are both normalized probability distributions
over the measurement outcomes of the second device, and can be compared
via the Bhattacharyya coefficient 
\begin{equation}
V_{\left(1\right)2}=\min_{\theta_{1},\theta_{2}}\sum_{b=x,y}\sqrt{\mathcal{P}_{b}\left(\theta_{2}\right)\mathcal{P}_{b}^{'}\left(\theta_{1},\theta_{2}\right)},
\end{equation}
which we minimize over the polarization angles, as we did with for
criterion based on LGIs. Note that whenever the probability distributions
are equal, $V_{(1)2}=1$ and the measurement on the first device has
no effect on the second. If that's not the case, then $V_{(1)2}<1$.
When three consecutive measurements are considered, then one also
needs to consider the joint probability distribution of two devices,
without and with the presence of a previous measurement; these are
given, respectively, by\begin{subequations} 
\begin{align}
\mathcal{P}_{bc}\left(\theta_{2},\theta_{3}\right) & =\frac{P_{bc}\left(\theta_{2},\theta_{3}\right)}{\sum_{b,c}P_{bc}\left(\theta_{2},\theta_{3}\right)},\\
\mathcal{P}_{bc}^{'}\left(\theta_{1},\theta_{2},\theta_{3}\right) & =\frac{\sum_{a}P_{abc}\left(\theta_{1},\theta_{2},\theta_{3}\right)}{\sum_{abc}P_{abc}\left(\theta_{1},\theta_{2},\theta_{3}\right)},
\end{align}
\end{subequations} and are again compared via the minimized Bhattacharyya
coefficient

\begin{equation}
V_{\left(1\right)23}=\min_{\theta_{1},\theta_{2},\theta_{3}}\sum_{b,c}\sqrt{\mathcal{P}_{bc}\left(\theta_{2},\theta_{3}\right)\mathcal{P}_{bc}^{'}\left(\theta_{1},\theta_{2},\theta_{3}\right)}.
\end{equation}
In all cases NSIT requires $V=1$ \cite{Kofler13}, so that $V<1$
witnesses a violation, and hence absence of macroscopic realism in
favor of quantum mechanics.

\section{Results}

In this section we summarize the main results found through an extensive
numerical analysis based on the expressions provided in the appendix.
We present them in two subsections. We start with $\mathcal{S}$ measurements,
and then follow with $\mathcal{B}$ and $\mathcal{F}$ measurements.
We discuss the results assuming $N\geq M$ for concreteness, but the
results are invariant under the exchange $N\leftrightarrow M$.

\subsection{$\mathcal{S}$ measurements}

A common trait to $\mathcal{S}_{\omega}$-type measurements is that
the results do not depend on the reflectivity $r$ for any choice
of $\omega$. In particular, the reflectivity appears only as a prefactor
in the different absolute probabilities $P_{a}$, $P_{ab}$, and $P_{abc}$,
which disappears once we consider normalized objects such as correlations
$C_{pq}$ or probability distributions $\mathcal{P}_{a}$, $\mathcal{P}_{a}^{'}$,
$\mathcal{P}_{ab}$, and $\mathcal{P}_{ab}^{'}$.

Let us first consider $\mathcal{S}_{1}$ measurements. It turns out
that there is a single Fock state that violates the LGI, namely state
$\left\vert 1,1\right\rangle $, for which $K$ is shown in Fig. \ref{Fig2}(a)
as a function of $\theta_{1}$ and $\theta_{2}$ for $\theta_{3}=\pi/2$.
Notice that the violation is maximal as $K_{\max}=3$. Regarding NSIT,
it turns out that $V_{\left(1\right)2}=V_{\left(1\right)23}=1$ for
all angles $\left(\theta_{1},\theta_{2},\theta_{3}\right)$ and for
all Fock states $\left\vert N,M\right\rangle $. Hence the violation
of the LGI by the state $\left\vert 1,1\right\rangle $ is not captured
by $V_{\left(1\right)2}$ (note that for state $\left\vert 1,1\right\rangle $
it does not make any sense to think about $V_{\left(1\right)23}$
as there are not enough photons for three detections).

The results are very different for $\mathcal{S}_{\omega>1}$ measurements.
In this case, there are an infinite number of states that violate
both the LGIs and NSIT for each $\omega$. Let us first consider LGIs.
In Fig. \ref{Fig2}(b) we show $K_{\max}$ as a function of $N$ and
$M$ for $\omega=2$. It can be appreciated that states with $1\lesssim M<M_{\max}\sim0.4N$
have $K_{\mathrm{max}}>1$. Hence the domain of states that violate
the LGI in $\mathcal{S}_{\omega>1}$ measurements is not bounded for
large values of $N$. For a given $N$ we have found by inspection
that the states with $M\sim N/6$ are the ones exhibiting a larger
violation, with a $K_{\max}$ quickly arriving to an asymptotic value
as $N$ increases, as shown in Fig. \ref{Fig4}(a) for several values
of $\omega$. Contrarily, for states with $M>M_{\max}$ there is an
upper value of $N$ beyond which the violation of the LGI disappears,
which is illustrated for states $\left\vert N,N\right\rangle $ in
Fig. \ref{Fig4}(b). This last result suggests that superpositions
of $\left\vert N,N\right\rangle $ states could violate LGIs only
provided $N$ is not too large, which could be illustrated with the
experimentally accessible two-mode squeezed vacuum states.

\begin{figure}
\includegraphics[width=0.9\columnwidth]{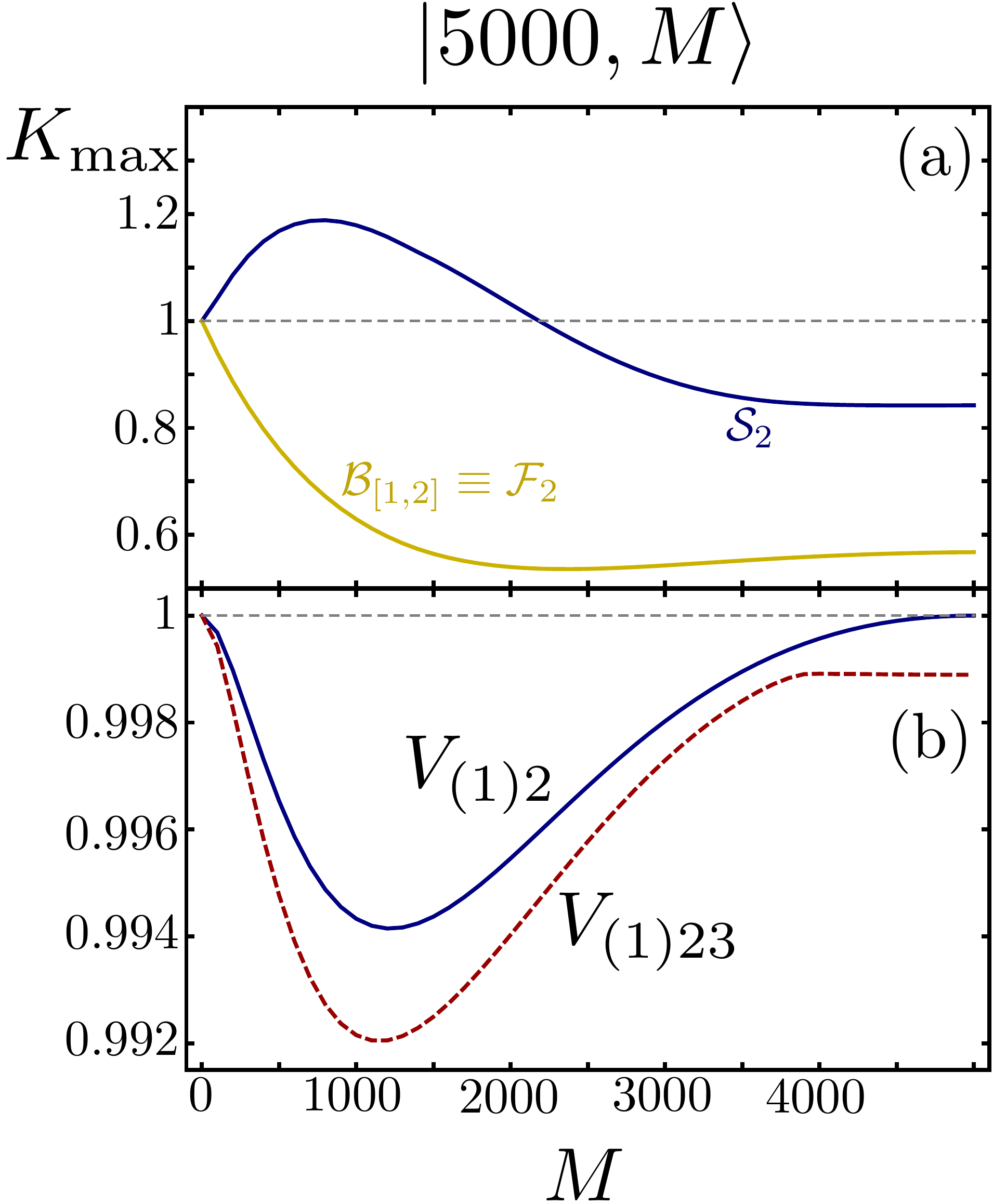}\caption{(a) $K_{\text{max}}$ for states $|5000,M\rangle$ as function of
$M$ for two types of measurements, $\mathcal{S}_{2}$ and $\mathcal{F}_{2}$
(with $r=0.1$ for the last one). (b) Bhattacharyya coefficients $V_{(1)2}$
and $V_{(1)23}$ for the same states as a function of $M$ subject
to $\mathcal{S}_{2}$ measurements.\label{Fig5}}
\end{figure}

Let us now move to NSIT. We have found $V_{\left(1\right)2}=1$ for
all states $\left\vert N,N\right\rangle $ and $\left\vert N,0\right\rangle $,
independently of $N$ and the choice of angles $\left(\theta_{1},\theta_{2}\right)$.
For the rest of Fock states, $V_{\left(1\right)2}$ is smaller than
one for small photon numbers, but rapidly grows towards an asymptotic
value close to one, which is typically reached for photon numbers
around $100$. The behavior of $V_{\left(1\right)23}$ is different,
as only states $\left\vert N,0\right\rangle $ have $V_{\left(1\right)23}=1$.

In Figs. \ref{Fig5} we illustrate this conclusions by considering
the states $\left\vert N=5000,M\right\rangle $, plotting different
quantities as a function of $M$ for $\mathcal{S}_{2}$ measurements.
We show $K_{\max}$ in Fig. \ref{Fig5}(a), which shows a maximum
at $M\sim800\sim5000/6$ as expected. In Fig. \ref{Fig5}(b) we show
$V_{\left(1\right)2}$ and $V_{\left(1\right)23}$, where we can appreciate
that $V_{\left(1\right)2}=1$ for $M=0$ and $M=5000$, while $V_{\left(1\right)23}=1$
only for $M=0$, both quantities being smaller than one for any other
$M$.

In summary, using $\mathcal{S}_{\omega>1}$ measurements both NSIT
and LGIs are violated by an infinite number of states, the violation
being larger for larger $\omega$, consistent with the fact that the
measurement becomes increasingly selective. The exceptions are completely
polarized states (those with $M=0$), for which no violation of macroscopic
realism is found, as expected from the fact that our photon subtraction
scheme cannot change their polarization state.

Let us remark that, while it might seem surprising the abrupt change
in the the number of states which violate macrorealism when moving
from $\mathcal{S}_{1}$ to $\mathcal{S}_{\omega>1}$ measurements,
these are indeed two very different types of measurements. In the
first case, one is really post-selecting to the most likely events
(reflection of a single photon), while in the second case, one post-selects
to increasingly unlikely events (reflection of many photons). Hence,
compared to $\mathcal{S}_{1}$ measurements, $\mathcal{S}_{\omega>1}$
measurements are extremely selective, and therefore, extremely invasive.

\subsection{$\mathcal{B}$ and $\mathcal{F}$ measurements}

In contrast to $\mathcal{S}$ measurements, for the $\mathcal{B}$
and $\mathcal{F}$ types the results depend on the reflectivity, as
we discuss next.

As an initial example, in Fig. \ref{Fig5}(a) we consider a $\mathcal{B}_{[1,2]}$
measurement (which is as well an $\mathcal{F}_{2}$ measurement),
showing how the violations of the LGI which we found for $\mathcal{S}_{2}$
measurements are completely smeared off when single photon detections
are also considered.

\begin{figure}[b]
\includegraphics[width=1\columnwidth]{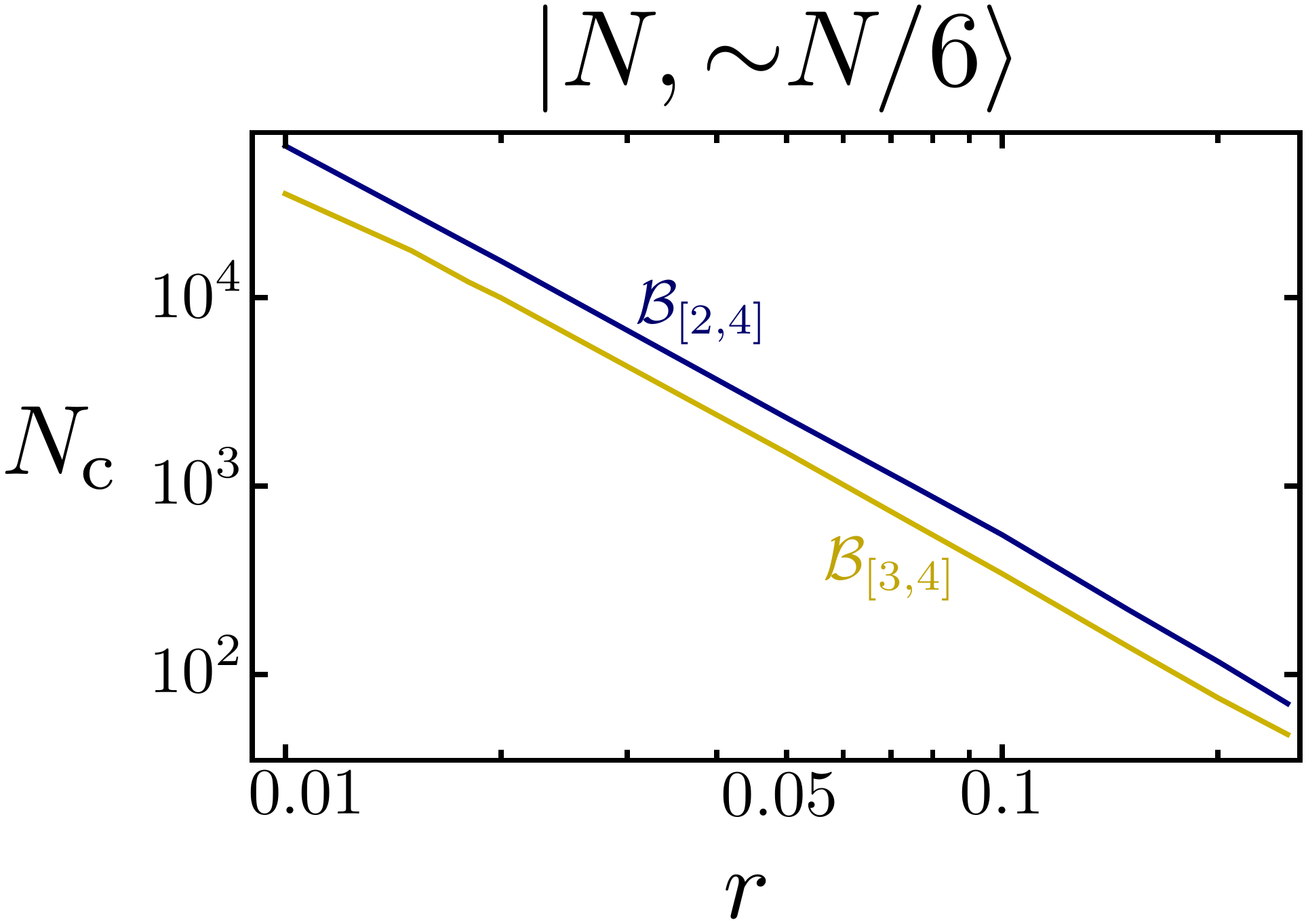}\caption{Critical value $N_{\text{c}}$ of the photon number above which violations
of LGI disappear as a function of the reflectivity $r$ for states
of the type $|N,\sim N/6\rangle$ and the types of $\mathcal{B}$
measurements indicated in the figure. Notice the logarithmic scale
in both axes. \label{Fig6}}
\end{figure}

\begin{figure}[t]
\includegraphics[width=1\columnwidth]{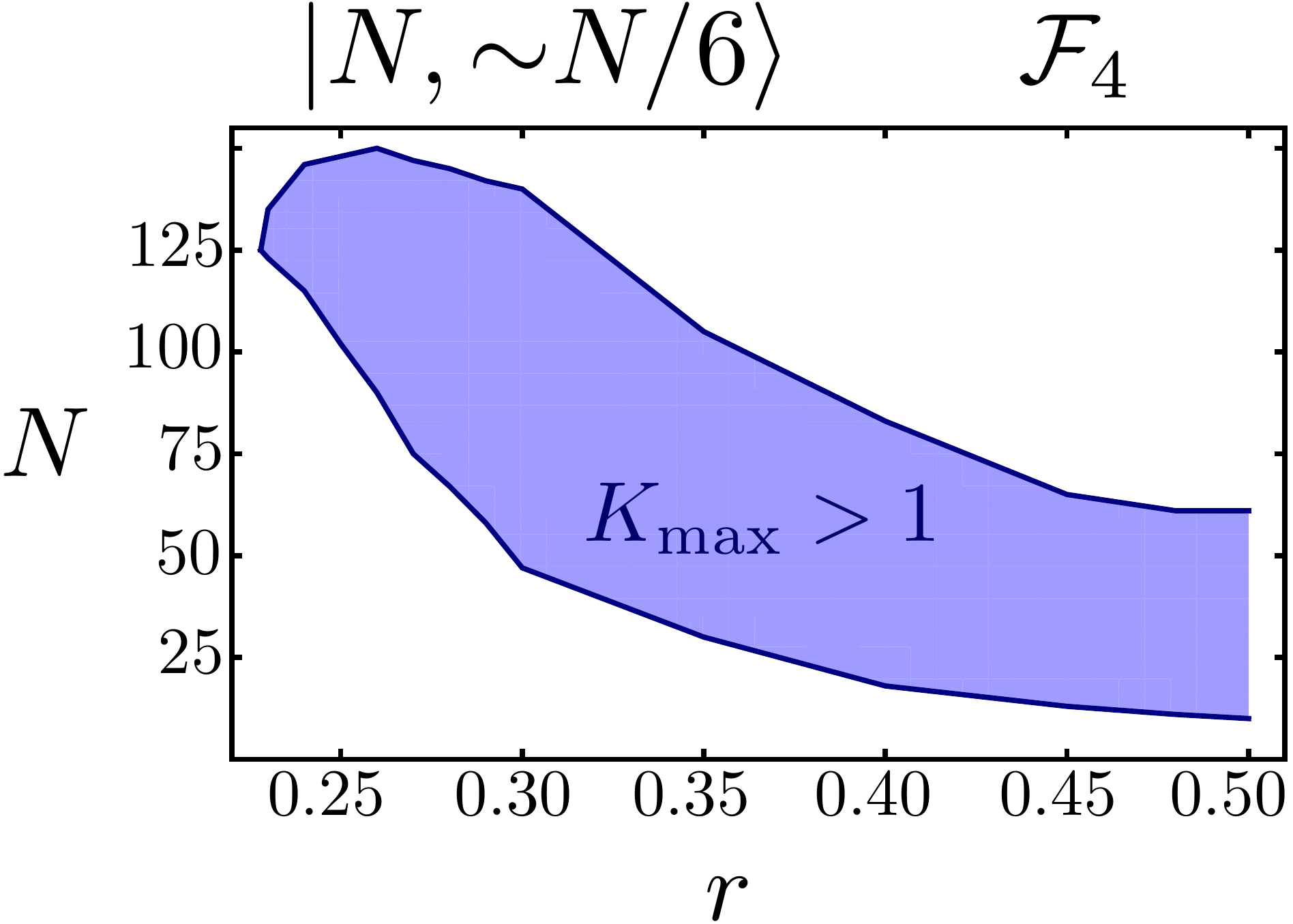}\caption{Region of violation of the LGI in the $(N,r)$ parameter space for
states of the $|N,\sim N/6\rangle$ type and $\mathcal{F}_{4}$ measurements.
\label{Fig7}}
\end{figure}

Let us now consider more complex measurements with $\omega_{\max}=4$
and study the effect of how selective the measurement is, focusing
on states with $M\sim N/6$, which we have found to be the ones violating
the LGI the strongest for $\mathcal{S}_{\omega>1}$ measurements.
For $\mathcal{S}_{4}$ measurements we saw that $K_{\mathrm{max}}$
reaches an asymptotic value above 1 as the photon number $N$ is increased.
In contrast, $\mathcal{B}_{[1<\omega_{\mathrm{min}}<4,4]}$ measurements
do not show such an asymptote, and indeed $K_{\mathrm{max}}$ becomes
smaller than 1 above some critical photon number $N_{\mathrm{c}}$.
In Fig. \ref{Fig6} we represent $N_{\mathrm{c}}$ as a function of
the reflectivity $r$ for $\mathcal{B}_{[3,4]}$ and $\mathcal{B}_{[2,4]}$
measurements, showing that coarse graining works against the violation
of macrorealism, that is, $N_{\mathrm{c}}$ is smaller the smaller
$\omega_{\mathrm{min}}$ is. The results are similar for $\mathcal{F}_{4}$
measurements, that is, there is a critical value of $N$ beyond which
there is no violation of the LGI, but in such case there are also
certain values of $N$ and the reflectivity below which $K_{\mathrm{max}}<1$.
We illustrate this in Fig. \ref{Fig7}, where the condition $K_{\max}>1$
is shown to lead to a closed domain in the space of parameters $(r,N)$.

\begin{figure}[t]
\includegraphics[width=1\columnwidth]{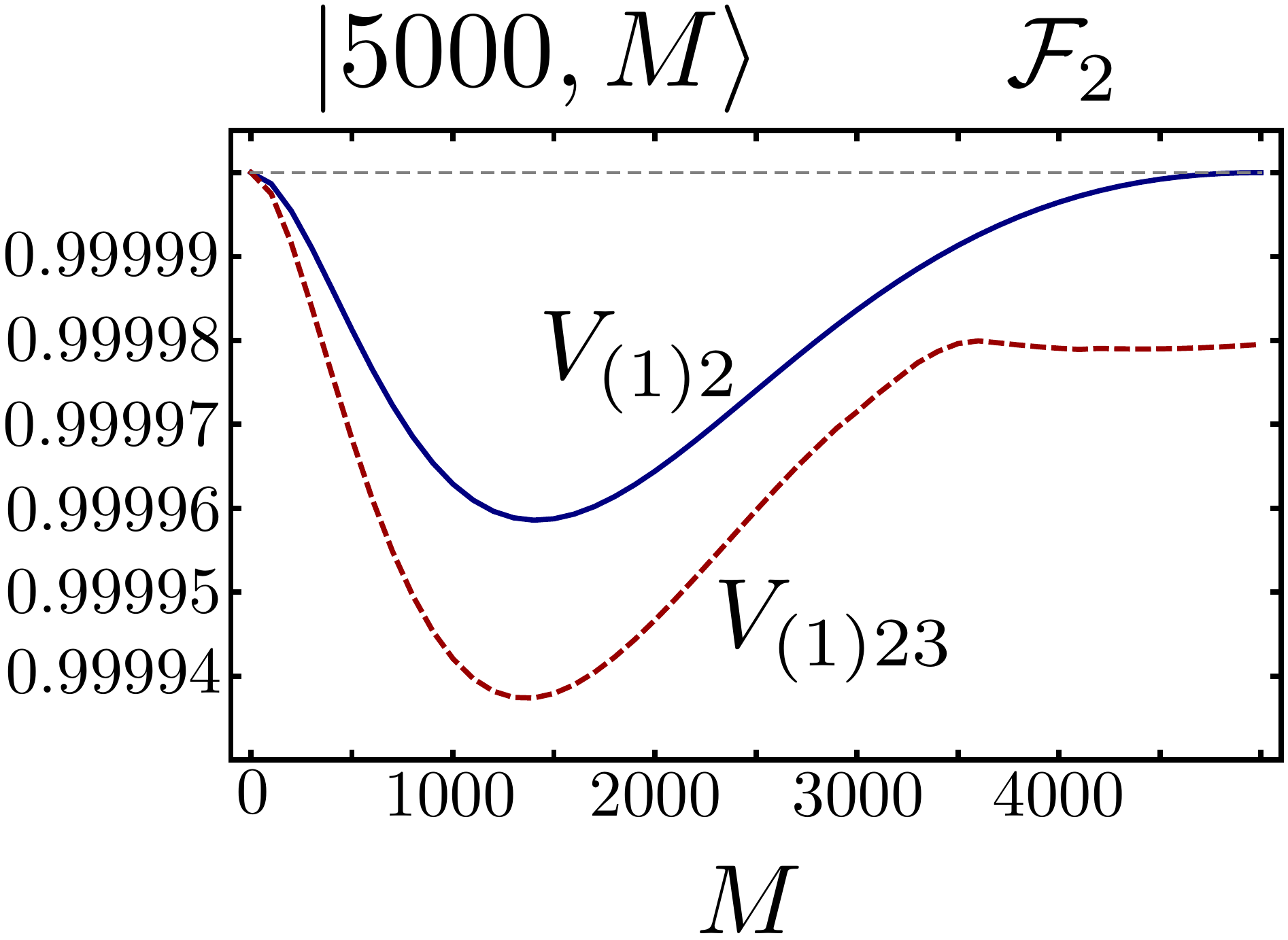}\caption{Bhattacharyya coefficients $V_{(1)2}$ and $V_{(1)23}$ as a function
of $M$ for states $|5000,M\rangle$ and $\mathcal{F}_{2}$ measurement.
We set $r=0.01$. \label{Fig8}}
\end{figure}

As for NSIT, our numerical analysis shows that it is violated for
all states except $\left\vert N,0\right\rangle $, but the violation
is weakened as the range $[\omega_{\mathrm{min}},\omega_{\mathrm{max}}]$
is increased or the reflectivity $r$ is reduced. As an example, in
Fig. \ref{Fig8} we show $V_{\left(1\right)2}$ and $V_{\left(1\right)23}$
as a function of $M$ for states $\left\vert N=5000,M\right\rangle $
and a $\mathcal{B}_{[1,2]}$ (which, we remark again, is an $\mathcal{F}_{2}$
measurement as well). The shape of the curves is essentially the same
as for $\mathcal{S}_{2}$ measurements, see Fig. 5(b), but their peak
values are much closer to one, and hence the violation of macrorealism
is weakened.

\section{Discussion and conclusions}

Let us finally offer some conclusions that can be drawn from the results
presented above and comment on possible future work.

Probably the most interesting question that our results might give
an answer to is: has the polarization of light a macrorealistic character?
In the setup we studied, the answer seems to be positive, because
violations of macrorealism are weakened as the invasiveness of the
measurements is reduced, that is, as we approach the condition $(N+M)r^{2}\ll1<\omega_{\mathrm{max}}$.
This conclusion is in agreement with the previous experimental analysis
that we commented on in the introduction \cite{Goggin11}.

A complementary question that our work answers as well is: does quantum
mechanics allow for a truly non-invasive way of measuring the polarization
of light? Recall that our setup is indeed capable of reconstructing
the statistics of the Stokes parameters, and hence the polarization
state. Hence, we can conclude that truly non-invasive polarization
measurements in the macroscopic domain occur only for $\mathcal{S}_{1}$
measurements (i.e., removing single photons), as NSIT is never violated,
while LGIs are solely violated by the state $|1,1\rangle$, certainly
not a macroscopic state. Any other type of measurement subtracting
more photons disturbs the system enough as to have violations of macrorealism.
However, we have also found that coarse-graining the measurements
and increasing the system size (input number of photons) tends to
make the violations disappear, so that polarization measurements may
be regarded as asymptotically non-invasive in this limit as well.

Note that these conclusions are drawn both from the analysis of LGIs
and NSIT. Hence, even though the conditions for macrorealism based
on these criteria are inequivalent in general (as explained above,
NSIT provides stronger conditions), in our setup both can be used
interchangeably. Moreover, if instead of a sharp $V=1$ macrorealism
condition for the Bhattacharyya coefficients, one allows for an inequality
type condition $V>V_{0}$ (with $V_{0}$ properly chosen for each
type of measurement), we have checked that the NSIT conditions reproduce
the results found with LGIs even quantitatively.

An interesting outlook that draws from our work is the analysis of
other types of input states, particularly those which make use of
the superposition principle. For example, considering GHZ states of
the form $\left\vert N,0\right\rangle +\left\vert 0,N\right\rangle $,
quantum mechanics predicts that even coarse-grained measurements would
lead to a violation of macrorealism in certain setups \cite{Kofler08,Jeong09}.
It will be interesting to analyze whether this is also the case in
our setup, even when the measurements can be considered non-invasive,
opening then the possibility to rule out either macrorealism or quantum
mechanics for the polarization degree of freedom of light.

Let us finally comment on a different type of input states that we
have considered, which have the form $\left\vert \alpha\right\rangle \otimes\left\vert N\right\rangle $,
that is, the mode polarized along the $x$ direction is in a coherent
state of amplitude $\alpha$, while the one along the $y$ direction
is in a Fock state with $N$ photons. This state is interesting because
it is a combination of a classical and a quantum state. Preliminary
results show that, for $\mathcal{S}$ measurements, the states that
maximally violate LGIs have $|\alpha|^{2}\sim N/6$, which is consistent
with we showed in the previous section. However, in this case the
dependence of $K_{\max}$ on the photon number $|\alpha|^{2}$ is
not as strong as it was with Fock states. Moreover, the domain of
angles where the violation occurred shrinks as $|\alpha|^{2}$ grows,
in contrast with what happens for the Fock state $\left\vert N,\sim N/6\right\rangle $
for which the domain of violation in the angle space is $N$-independent.
Hence, increasing the ``classicality'' of the state works agains
violations of macrorealism, as expected.

\textbf{Acknowledgements}. This work has benefited from discussions
with our colleagues Germán J. de Valcárcel and J. Ignacio Cirac. We
acknowledge financial support from the Spanish government (Ministerio
de Economía y Competitividad) and the European Union FEDER through
project FIS2014-60715-P. CN-B acknowledges funding from the Alexander
von Humboldt Foundation through their Fellowship for Postdoctoral
Researchers.

\appendix

\section{Determination of probabilities}

In this appendix we derive the expressions for the conditional probabilities
appearing in the LGI and the NSIT conditions. We proceed by determining
the un-normalized state of the system after obtaining a given outcome
(photon counts) on each of the measurement ports. The norm of this
state provides the probability of those particular outcomes.

\subsection{First measurement}

At each detection port, the pulses undergo a series of transformations:
first, they mixed in a beam splitter with a vacuum state, then the
polarization of the reflected pulse is rotated by an angle $\theta$,
and finally photon counters measure a given number of photons in the
$x$ and $y$ polarizations of this pulse, so that the transmitted
state gets projected to the corresponding outcome (although physically
different, this is equivalent to the effect of the calcite crystal
of Fig. \ref{Fig1}, in which the two orthogonal linear polarizations
along some angle $\theta$ are separated and photon-counted). Let
us find the post-selected state now.

Consider an input state 
\begin{equation}
\left\vert N,M\right\rangle =\frac{1}{\sqrt{N!M!}}\left(\hat{a}_{x}^{\dagger}\right)^{N}\left(\hat{a}_{y}^{\dagger}\right)^{M}\left\vert 0\right\rangle ,
\end{equation}
where $\hat{a}_{c}^{\dagger}$ are creation operators for $c$-polarized
photons. In the following we will denote by $|0\rangle$ the vacuum
state of whatever number of modes we are dealing with. The action
of the beam splitter is described \cite{NavarreteBook15} by the unitary
operator $\hat{B}(r)=\exp(\alpha\hat{a}_{x}^{\dagger}\hat{b}_{x}+\alpha\hat{a}_{y}\hat{b}_{y}^{\dagger}-\mathrm{H.c.})$,
where $r=\cos\alpha$ and $t=\sin\alpha$ are the reflectivity and
transmissivity of the beam splitter, and $\hat{b}_{c}$ are the annihilation
operators for the second port. Applying this operator to the input
state, including the vacuum state for the modes of the second port,
we obtain 
\begin{align}
\left\vert \psi_{1}\right\rangle  & =\hat{B}(r)\left\vert N,M\right\rangle \otimes|0\rangle\\
 & =\frac{1}{\sqrt{N!M!}}\left(t\hat{a}_{x}^{\dagger}+r\hat{b}_{x}^{\dagger}\right)^{N}\left(t\hat{a}_{y}^{\dagger}+r\hat{b}_{y}^{\dagger}\right)^{M}\left\vert 0\right\rangle \nonumber \\
 & =\frac{t^{N+M}}{\sqrt{N!M!}}\sum_{n=0}^{N}\sum_{m=0}^{M}\left(\frac{r}{t}\right)^{n+m}\tbinom{N}{n}\tbinom{M}{m}\nonumber \\
 & \hspace{1.5cm}\times\left(\hat{a}_{x}^{\dagger}\right)^{N-n}\left(\hat{a}_{y}^{\dagger}\right)^{M-m}\left(\hat{b}_{x}^{\dagger}\right)^{n}\left(\hat{b}_{y}^{\dagger}\right)^{m}\left\vert 0\right\rangle .\nonumber 
\end{align}
The rotation of the polarization of the reflected modes is described
by the unitary operator $\hat{R}(\theta)=\exp(\theta b_{x}^{\dagger}\hat{b}_{y}-\mathrm{H.c.})$.
We will denote the shorthand notation $s=\sin\theta$ and $c=\cos\theta$
in what follows. Applying this transformation to the previous state
we obtain 
\begin{align}
\left\vert \psi_{2}\right\rangle  & =\hat{R}(\theta)\left\vert \psi_{1}\right\rangle \\
= & \frac{t^{N+M}}{\sqrt{N!M!}}\sum_{n=0}^{N}\sum_{m=0}^{M}\left(\frac{r}{t}\right)^{n+m}\tbinom{N}{n}\tbinom{M}{m}\left(\hat{a}_{x}^{\dagger}\right)^{N-n}\nonumber \\
 & \hspace{1cm}\times\left(\hat{a}_{y}^{\dagger}\right)^{M-m}\underset{|0\rangle\otimes\left\vert \mathcal{Z}_{nm}(\theta)\right\rangle }{\underbrace{\left(c\hat{b}_{x}^{\dagger}-s\hat{b}_{y}^{\dagger}\right)^{n}\left(s\hat{b}_{x}^{\dagger}+c\hat{b}_{y}^{\dagger}\right)^{m}\left\vert 0\right\rangle }},\nonumber 
\end{align}
Suppose that the photon-counters detect $\omega_{x}$ and $\omega_{y}$
photons in the corresponding mode. Defining the Fock-basis projector
$\hat{\Pi}_{\omega_{x}\omega_{y}}=\left\vert \omega_{x}\right\rangle \left\langle \omega_{x}\right\vert \otimes\left\vert \omega_{y}\right\rangle \left\langle \omega_{y}\right\vert $,
the state of the transmitted modes is finally transformed \cite{NavarreteBook15}
into the (un-normalized) state $\mathrm{tr}_{b}\{(\hat{I}\otimes\hat{\Pi}_{\omega_{x}\omega_{y}})\left\vert \psi_{2}\right\rangle \left\langle \psi_{2}\right\vert \}=\left\vert \varphi\right\rangle \left\langle \varphi\right\vert $,
with 
\begin{align}
\left\vert \varphi\right\rangle  & =\frac{t^{N+M}}{\sqrt{N!M!}}\sum_{n=0}^{N}\sum_{m=0}^{M}\left(\frac{r}{t}\right)^{n+m}\tbinom{N}{n}\tbinom{M}{m}\\
 & \hspace{1.5cm}\times\left(\hat{a}_{x}^{\dagger}\right)^{N-n}\left(\hat{a}_{y}^{\dagger}\right)^{M-m}\langle\omega_{x},\omega_{y}\left\vert \mathcal{Z}_{nm}(\theta)\right\rangle \left\vert 0\right\rangle .\nonumber 
\end{align}
Let us evaluate $\langle\omega_{x},\omega_{y}\left\vert \mathcal{Z}_{nm}(\theta)\right\rangle $
separately, for which we first rewrite 
\begin{align}
\left\vert \mathcal{Z}_{nm}(\theta)\right\rangle  & =\sum_{i=0}^{n}\sum_{j=0}^{m}\left(-1\right)^{i}\tbinom{n}{i}\tbinom{m}{j}c_{\theta}^{n-i+j}s_{\theta}^{m+i-j}\\
 & \hspace{2.5cm}\times\left(\hat{b}_{x}^{\dagger}\right)^{n+m-i-j}\left(\hat{b}_{y}^{\dagger}\right)^{i+j}\left\vert 0\right\rangle ,\nonumber 
\end{align}
and then calculate 
\begin{align}
\langle\omega_{x},\omega_{y}\left\vert \mathcal{Z}_{nm}(\theta)\right\rangle  & =\sum_{i=0}^{n}\sum_{j=0}^{m}\left(-1\right)^{i}\tbinom{n}{i}\tbinom{m}{j}c^{n-i+j}s^{m+i-j}\nonumber \\
 & \times\sqrt{\left(n+m-i-j\right)!\left(i+j\right)!}\\
 & \underset{\delta_{n+m-i-j,\omega_{x}}}{\times\underbrace{\left\langle \omega_{x}\right.\left\vert n+m-i-j\right\rangle }}\underset{\delta_{i+j,\omega_{y}}}{\underbrace{\left\langle \omega_{y}\right.\left\vert i+j\right\rangle }}\nonumber \\
= & \sum_{i=i_{\min}}^{i_{\max}}\left(-1\right)^{i}\tbinom{n}{i}\tbinom{m}{\omega_{y}-i}\nonumber \\
 & \times c^{\omega_{y}+n-2i}s^{\omega_{x}-n+2i}\sqrt{\omega_{x}!\omega_{y}!}\nonumber 
\end{align}
with limits $i_{\min}=\max\{0,n-\omega_{x}\}$ and $\ i_{\max}=\min\left\{ \omega_{y},n\right\} $
which are provided by the existence condition of the elements in the
sum. Then, the post-selected un-normalized state after the first measurement
device can be written as 
\begin{align}
\left\vert \varphi\right\rangle  & =\frac{t^{N+M}}{\sqrt{N!M!}}\left(\frac{r}{t}\right)^{\omega_{x}+\omega_{y}}\sqrt{\omega_{x}!\omega_{y}!}\\
 & \times\sum_{n=\max\left\{ 0,\omega_{x}+\omega_{y}-M\right\} }^{\min\left\{ \omega_{x}+\omega_{y},N\right\} }\tbinom{N}{n}\tbinom{M}{\omega_{x}+\omega_{y}-n}\nonumber \\
 & \times\sum_{i=i_{\min}}^{i_{\max}}\left(-1\right)^{i}\tbinom{n}{i}\tbinom{\omega_{x}+\omega_{y}-n}{\omega_{y}-i}\nonumber \\
 & \times c^{\omega_{y}+n-2i}s^{\omega_{x}-n+2i}\left(\hat{a}_{x}^{\dagger}\right)^{N-n}\left(\hat{a}_{y}^{\dagger}\right)^{M-\omega_{x}-\omega_{y}+n}\left\vert 0\right\rangle ,\nonumber 
\end{align}
where again the limits in the summation are imposed by the existence
conditions of the terms in the sum.

The state above can be written in a clearer and more compact notation
as 
\begin{equation}
\left\vert \varphi\right\rangle =\sum_{n=n_{\text{min}}}^{n_{\text{max}}}\mathcal{A}_{n,\mathbf{w}_{1}}\left(N,M,\theta,r\right)\left\vert N-n,M+n-\omega_{1}\right\rangle ,\label{fi}
\end{equation}
with multi-index $\mathbf{w}_{1}=\left(\omega_{x},\omega_{y}\right)$,
limits $n_{\text{min}}=\max\left\{ 0,\omega_{1}-M\right\} $ and $n_{\text{max}}=\max\left\{ \omega_{1},N\right\} $,
where $\omega_{1}=\omega_{x}+\omega_{y}$ the number of detected photons,
and 
\begin{align}
 & \mathcal{A}_{n,\mathbf{w}_{1}}\left(N,M,\theta,r\right)=t^{N+M}\left(\frac{r}{t}\right)^{\omega_{1}}\tbinom{N}{n}\tbinom{M}{\omega_{1}-n}\\
 & \hspace{1.7cm}\times\sqrt{\frac{(N-n)!(M+n-\omega_{1})!\omega_{x}!\omega_{y}!}{N!M!}}\nonumber \\
 & \times\sum_{i=\max\{0,n-\omega_{x}\}}^{\min\left\{ \omega_{y},n\right\} }\left(-1\right)^{i}\tbinom{n}{i}\tbinom{\omega_{1}-n}{\omega_{y}-i}\cos^{\omega_{y}+n-2i}\sin^{\omega_{x}-n+2i}.\nonumber 
\end{align}
Note that although it is not explicitly denoted on its label, $\left\vert \varphi\right\rangle $
depends on all the relevant parameters $\left(\omega_{x},\omega_{y},N,M,\theta,r\right)$.

Being $\left\vert \varphi\right\rangle $ the post-selected un-normalized
state after $\omega_{x}$ and $\omega_{y}$ photons are detected,
its norm provides the probability $\bar{P}_{\mathbf{w}_{1}}$ of detecting
this number of photons at the first measurement 
\begin{equation}
\bar{P}_{\mathbf{w}_{1}}\left(N,M,\theta,r\right)=\sum_{n=n_{\text{min}}}^{n_{\text{max}}}\left\vert \mathcal{A}_{n,\mathbf{w}_{1}}\left(N,M,\theta,r\right)\right\vert ^{2}.\label{P1}
\end{equation}

From this expression, which is easily computed with the help of a
computer, it is simple to find the probability for an $a$-event in
the different types of measurements. For example, for an $\mathcal{S}_{\omega}$
measurement, we have $P_{x}=\bar{P}_{(\omega,0)}$, while for a $\mathcal{B}_{[\omega_{\text{min}},\omega_{\text{max}}]}$
we have $P_{x}=\sum_{\omega_{x},\omega_{y}=\omega_{\text{min}}}^{\omega_{\text{max}}}H(\omega_{x}-\omega_{y})\bar{P}_{(\omega_{x},\omega_{y})}$,
where $H(z)$ is the step function defined as 0 for $z\leq0$ and
1 for $z>0$. 

\subsection{Second measurement}

The post-selected state $\left\vert \varphi\right\rangle $ enters
a second measurement port with a different orientation $\theta^{\prime}$
of the calcite-crystal axes. We can easily derive the expression of
the post-selected state after counting $\omega_{x}^{\prime}$ and
$\omega_{y}^{\prime}$ photons in the detectors. To this aim, we just
note that since $\left\vert \varphi\right\rangle $ is written in
the Fock basis as a superposition of $\left\vert N^{\prime},M^{\prime}\right\rangle $
states, with $N^{\prime}=N-n$ and $M^{\prime}=M+n-\omega_{1}$, the
only thing we need is to find the transformation of these states.
It is clear that the transformed $\left\vert N^{\prime},M^{\prime}\right\rangle $
will have the same expression as (\ref{fi}), but replacing $N$,
$M$, and $\theta$ by $N^{\prime}$, $M^{\prime}$, and $\theta^{\prime}$,
respectively. Hence, the un-normalized transmitted state after the
second measurement can be written as 
\begin{align}
\left\vert \varphi^{\prime}\right\rangle  & =\sum_{n=n_{\text{min}}}^{n_{\text{max}}}\sum_{k=k_{\min}}^{k_{\max}}\mathcal{A}_{n,\mathbf{w}_{1}}\left(N,M,\theta,r\right)\\
 & \hspace{1.7cm}\times\mathcal{A}_{k,\mathbf{w}_{1}^{\prime}}\left(N-n,M+n-\omega_{1},\theta^{\prime},r\right)\nonumber \\
 & \hspace{1.7cm}\times\left\vert N-k-n,M+k+n-\omega_{12}\right\rangle ,\nonumber 
\end{align}
with $\mathbf{w}_{1}^{\prime}=\left(\omega_{x}^{\prime},\omega_{y}^{\prime}\right)$,
limits $k_{\min}=\max\left\{ 0,\omega_{12}-n-M\right\} $ and $k_{\max}=\min\left\{ \omega_{1}^{\prime},N-n\right\} $,
and where $\omega_{1}^{\prime}=\omega_{x}^{\prime}+\omega_{y}^{\prime}$
is the number of photons detected in the second measurement, while
$\omega_{12}=\omega_{x}+\omega_{y}+\omega_{x}^{\prime}+\omega_{y}^{\prime}$
is the total number of detected photons. Noticing that $k$ and $n$
appear in the states only through the combination $s=n+k$, it is
convenient to write $k=s-n$, and change the sum in $k$ by a sum
in $s$, easily arriving at 
\begin{equation}
\left\vert \varphi^{\prime}\right\rangle =\sum_{s=s_{\text{min}}}^{s_{\text{max}}}\mathcal{B}_{s,\mathbf{w}_{2}}\left(N,M,\theta,\theta^{\prime},r\right)\left\vert N-s,M+s-\omega_{12}\right\rangle ,
\end{equation}
with $\mathbf{w}_{2}=\left(\omega_{x},\omega_{y},\omega_{x}^{\prime},\omega_{y}^{\prime}\right)$,
limits $s_{\text{min}}=\max\left\{ 0,\omega_{12}-M\right\} $ and
$s_{\text{max}}=\min\left\{ \omega_{12},N\right\} $, and 
\begin{align}
\mathcal{B}_{s,\mathbf{w}_{2}}\left(N,M,\theta,\theta^{\prime},r\right)=\sum_{m=m_{\min}}^{m_{\max}}\mathcal{A}_{m,\mathbf{w}_{1}}\left(N,M,\theta,r\right)\\
\times\mathcal{A}_{s-m,\mathbf{w}_{1}^{\prime}}\left(N-m,M+m-\omega_{1},\theta^{\prime},r\right) & ,\nonumber 
\end{align}
with limits $m_{\min}=\max\left\{ 0,\omega_{1}-M,s-\omega_{1}^{\prime}\right\} $
and $m_{\max}=\min\left\{ \omega_{1},N,s\right\} .$

The un-normalized state $\left\vert \varphi^{\prime}\right\rangle $
is the post-selected state after $\omega_{x}$ and $\omega_{y}$ photons
are detected at the first measurement device, and $\omega_{x}^{\prime}$
and $\omega_{y}^{\prime}$ photons are detected at the second measurement
device. The corresponding probability $\bar{P}_{\mathbf{w}_{2}}$
of detecting this number of photons is then given by its norm, which
reads 
\begin{equation}
\bar{P}_{\mathbf{w}_{2}}\left(N,M,\theta,\theta^{\prime},r\right)=\sum_{s=s_{\text{min}}}^{s_{\text{max}}}\left\vert \mathcal{B}_{s,\mathbf{w}_{2}}\left(N,M,\theta,\theta^{\prime},r\right)\right\vert ^{2}.\label{P2}
\end{equation}

Once we have this bare conditional probabilities for photon counts,
we can find the probability $P_{ab}$ of an $ab$-event for the different
type of measurements. In the case of $\mathcal{S}_{\omega}$ measurements,
one easily writes $P_{xx}=\bar{P}_{(\omega,0,\omega,0)}$ and $P_{xy}=\bar{P}_{(\omega,0,0,\omega)}$,
for example.  In contrast, considering a more general $\mathcal{B}_{[\omega_{\text{min}},\omega_{\text{max}}]}$
measurement, we have
\begin{align}
P_{xx}=\\
 & \hspace{-0.7cm}\sum_{\omega_{x},\omega_{x}^{\prime}=\omega_{\min}}^{\omega_{\max}}\left(P_{(\omega_{x},0,\omega_{x}^{\prime},0)}+\sum_{\omega_{y}=\omega_{\min}}^{\omega_{x}-1}\sum_{\omega_{y}^{\prime}=\omega_{\min}}^{\omega_{x}^{\prime}-1}P_{(\omega_{x},\omega_{y},\omega_{x}^{\prime},\omega_{y}^{\prime})}\right.\nonumber \\
 & \hspace{1cm}+(1-\delta_{\omega_{x}^{\prime}\omega_{\text{min}}})\sum_{\omega_{y}^{\prime}=\omega_{\min}}^{\omega_{x}^{\prime}-1}P_{(\omega_{x},0,\omega_{x}^{\prime},\omega_{y}^{\prime})}\nonumber \\
 & \hspace{1cm}\left.+(1-\delta_{\omega_{x}\omega_{\text{min}}})\sum_{\omega_{y}=\omega_{\min}}^{\omega_{x}-1}P_{(\omega_{x},\omega_{y},\omega_{x}^{\prime},0)}\right),\nonumber 
\end{align}
and 
\begin{align}
P_{xy} & =\\
 & \hspace{-0.7cm}\sum_{\omega_{x},\omega_{y}^{\prime}=\omega_{\min}}^{\omega_{\max}}\left(P_{(\omega_{x},0,0,\omega_{y}^{\prime})}+\sum_{\omega_{y}=\omega_{\min}}^{\omega_{x}-1}\sum_{\omega_{x}^{\prime}=\omega_{\min}}^{\omega_{y}^{\prime}-1}P_{(\omega_{x},\omega_{y},\omega_{x}^{\prime},\omega_{y}^{\prime})}\right)\nonumber \\
 & \hspace{1cm}+\sum_{\omega_{x}=\omega_{\min}}^{\omega_{\max}}\sum_{\omega_{y}^{\prime}=\omega_{\min}+1}^{\omega_{\max}}\sum_{\omega_{x}^{\prime}=\omega_{\min}}^{\omega_{y}^{\prime}-1}P_{(\omega_{x},0,\omega_{x}^{\prime},\omega_{y}^{\prime})}\nonumber \\
 & \hspace{1cm}+\sum_{\omega_{y}^{\prime}=\omega_{\min}}^{\omega_{\max}}\sum_{\omega_{y}=\omega_{\min}+1}^{\omega_{\max}}\sum_{\omega_{x}=\omega_{\min}}^{\omega_{y}-1}P_{(\omega_{x},\omega_{y},0,\omega_{y}^{\prime})}.\nonumber 
\end{align}

Similar expressions can be written for the other types of $ab$-events.
These together with the probabilities for the $a$-events of the previous
section are all we need to evaluate correlation functions and LGIs.

\subsection{Third measurement}

In order to study no signaling in time, we need to consider a third
measurement (characterized by the angle $\theta^{\prime\prime}$ and
a number of detected photons $\omega_{x}^{\prime\prime}$ and $\omega_{y}^{\prime\prime}$).
The derivation follows the same lines we have seen above and the result
reads 
\begin{align}
\left\vert \varphi^{\prime\prime}\right\rangle  & =\sum_{t=t_{\text{min}}}^{t_{\text{max}}}\mathcal{C}_{t,\mathbf{w}_{3}}\left(N,M,\theta,\theta^{\prime},\theta^{\prime\prime},r\right)\\
 & \hspace{1.5cm}\times\left\vert N-s,M+s-\omega_{123}\right\rangle ,\nonumber 
\end{align}
with $\mathbf{w}_{3}=\left(\omega_{x},\omega_{y},\omega_{x}^{\prime},\omega_{y}^{\prime},\omega_{x}^{\prime\prime},\omega_{y}^{\prime\prime}\right)$,
limits $t_{\text{min}}=\max\left\{ 0,\omega_{123}-M\right\} $ and
$t_{\text{max}}=\min\left\{ \omega_{123},N\right\} $, $\omega_{123}=\omega_{x}+\omega_{y}+\omega_{x}^{\prime}+\omega_{y}^{\prime}+\omega_{x}^{\prime\prime}+\omega_{y}^{\prime\prime}$
the total number of detected photons, and 
\begin{align}
\mathcal{C}_{t,\mathbf{w}_{3}}\left(N,M,\theta,\theta^{\prime},\theta^{\prime\prime},r\right)=\sum_{l=l_{\min}}^{l_{\max}}\mathcal{B}_{n,\mathbf{w}_{2}}\left(N,M,\theta,\theta^{\prime},r\right)\\
\times\mathcal{A}_{t-l,\mathbf{w}_{1}^{\prime\prime}}\left(N-l,M+l-\omega_{12},\theta^{\prime\prime},r\right),\nonumber 
\end{align}
where $\mathbf{w}_{1}^{\prime\prime}=(\omega_{x}^{\prime\prime},\omega_{y}^{\prime\prime})$,
the limits are $l_{\min}=\max\left\{ 0,\omega_{12}-M,t-\omega_{1}^{\prime\prime}\right\} $
and $l_{\max}=\min\left\{ \omega_{12},N,t\right\} $, and $\omega_{1}^{\prime\prime}=\omega_{x}^{\prime\prime}+\omega_{y}^{\prime\prime}$
are the photons detected in the third measurement port.

Finally, we evaluate the probability $\bar{P}_{\mathbf{w}_{3}}$ of
measuring the sequence $\mathbf{w}_{3}$ of photon numbers as 
\begin{equation}
\bar{P}_{\mathbf{w}_{3}}\left(N,M,\theta,\theta^{\prime},\theta^{\prime\prime},r\right)=\sum_{t=t_{\text{min}}}^{t_{\text{max}}}\left\vert \mathcal{C}_{t,\mathbf{w}_{3}}\left(N,M,\theta,\theta^{\prime},\theta^{\prime\prime},r\right)\right\vert ^{2}.\label{P3}
\end{equation}

From this expression, we can evaluate the probability $P_{abc}$ for
an $abc$-event for the different types of measurements. We don't
write the general expressions here because they are too lengthy in
the general case, but they are trivially found following the same
lines as with $ab$-events and $a$-events.

\end{document}